# Classical Mathematics for a Constructive World

BY RUSSELL O'CONNOR

Department of Computing and Software, McMaster University
Microsoft Research - INRIA Joint Centre

*Email:* `roconn@mcmaster.ca`

**Abstract**

Interactive theorem provers based on dependent type theory have the flexibility to support both constructive and classical reasoning. Constructive reasoning is supported natively by dependent type theory and classical reasoning is typically supported by adding additional non-constructive axioms. However, there is another perspective that views constructive logic as an extension of classical logic. This paper will illustrate how classical reasoning can be supported in a practical manner inside dependent type theory without additional axioms. We will see several examples of how classical results can be applied to constructive mathematics. Finally, we will see how to extend this perspective from logic to mathematics by representing classical function spaces using a weak value monad.

## 0 Licence

## 1 Introduction

The common view of constructive logic is that it is a restriction of classical logic where the law of the excluded middle and proof by contradiction are not allowed. Several software proof assistants are founded on constructive logic, including Coq [The Coq Development Team, 2009] and Agda [Norell, 2007], and they use the Curry-Howard isomorphism to interpret constructive deductions as programs and vice versa.

In accordance with this common view of constructive logic, these systems can be extended to classical logic by adding non-constructive axioms such as the law of excluded middle or proof by contradiction. When these axioms are added, all the constructive theorems can be used in classical proofs, since they are all still valid classical deductions. However, one cannot use classical proofs in constructive theorems since they rely on non-constructive axioms.

There is a high price to be paid by these non-constructive axioms. If one attempts to evaluate a classical deduction (viewed through the Curry-Howard isomorphism), the evaluation could get stuck at a non-constructive axiom because these axioms do not have a computational interpretation.

We will not be taking this common viewpoint in this paper. Instead, we will see constructive logic as an extension of classical logic. We will add two new logical connectives to classical logic: the constructive disjunction, written as $\varphi + \psi$, and the constructive existential, written as $\Sigma a\colon A.\, \varphi(a)$. Constructive theorems will make use of these new connectives while theorems living in the classical fragment will not. These constructive connectives are stronger than their classical counterparts in the sense that the following formulas hold:

$$(\varphi + \psi) \;\Rightarrow\; (\varphi \vee \psi) \tag{1}$$
$$(\Sigma a\colon A.\varphi(a)) \;\Rightarrow\; (\exists a\colon A.\varphi(a)) \tag{2}$$

This viewpoint allows us to freely mix classical and constructive theorems since a classical theorem is simply a special type of constructive theorem. In many cases, constructive results can be used in classical theorems by weakening them using the above theorems, but we will also see some examples where classical theorems are used in constructive proofs.

This alternative viewpoint is not original, but it is perhaps not widely known in the classical mathematics community. Even in constructive circles where it is known to some people, it is perhaps underappreciated. The purpose of this paper is to

1. Illustrate that this alternative viewpoint is a useful way to practice classical and constructive mathematics,

2. Illustrate that classical theorems, when seen this way, are useful when creating constructive proofs,

3. Extend this viewpoint from logic to mathematics by defining a classical function space as a special kind of constructive function space.

Before proceeding, it is worth remarking on my notation. Usually papers on constructive mathematics usurp the classical logical symbols, taking $\vee$ and $\exists$ to mean constructive disjunction and constructive existence respectively. I feel that this obscures the fact that the constructive result is stronger than the classical result. Also, taking the same symbols to have different meanings is a source of confusion and a potential source of conflict between the constructive and classical mathematics communities. Since our goal here is to unify and to bring classical mathematics into the world of constructive mathematics, I will be using new symbols for constructive logical operators. For constructively minded readers, please take careful note of the definitions of logical operators in Section 2.1.

Constructive mathematics also has new mathematical definitions that are analogous to common mathematical definitions but use these constructive logical operators, such as real numbers, continuity, etc. Since this paper is about constructive mathematics, it would be tedious to prefix "constructive" in front of every mathematical term. Instead, I will use unqualified terms, such as "real numbers", for the constructive definitions, and prefix "classical" when referring to "classical real numbers". I hope the reader will not find this too confusing.

## 2 Logic

My intention is to use dependent type theory viewed through the Curry-Howard isomorphism [Thompson, 1991] as my deduction system for constructive logic. However, to help ease the unfamiliar reader into the constructive world and to illustrate how constructive logic is really an extension of classical logic, we will slowly and informally unveil dependent type theory beginning first with the classical fragment of constructive logic.

### 2.1 Classical logic

Before defining the logic, we will define a typed term language. To get us started, we will assume we are given types for booleans $\mathbb{B}$, natural numbers $\mathbb{N}$, and higher-order functions, $A \Rightarrow B$, over these basic types.

The booleans come with two constructors: true, $\mathsf{T}\colon \mathbb{B}$, and false, $\mathsf{F}\colon \mathbb{B}$. There is an if-then-else function for elimination:
$$\mathrm{if}_A\colon \mathbb{B} \Rightarrow A \Rightarrow A \Rightarrow A$$
If-then-else comes with its standard reduction rules:
$$\mathrm{if}\,\mathsf{T}\,x\,y \;\rightsquigarrow\; x$$
$$\mathrm{if}\,\mathsf{F}\,x\,y \;\rightsquigarrow\; y$$
Using if-then-else, one can define any boolean operator.

The natural numbers come with constructors zero, $0\colon \mathbb{N}$, and successor, $\mathsf{S}\colon \mathbb{N} \Rightarrow \mathbb{N}$; and an eliminator for primitive recursion:
$$\mathrm{rec}_A\colon \mathbb{N} \Rightarrow A \Rightarrow (\mathbb{N} \Rightarrow A \Rightarrow A) \Rightarrow A$$

Primitive recursion comes with its standard reduction rules:

$$\operatorname{rec}_A 0\, z\, f \rightsquigarrow z$$
$$\operatorname{rec}_A (\mathsf{S} n)\, z\, f \rightsquigarrow f n (\operatorname{rec}_A n\, z\, f)$$

Higher-order functions will be defined by lambda expressions and comes with the standard reduction rule:

$$(\lambda x.e) y \rightsquigarrow e[x \mapsto y]$$

With primitive recursion, we can define standard operations over natural numbers such as $+$ and $\times$. We can also define equality between natural numbers:[1]

$$\begin{aligned} 0 =_\mathbb{N} 0 &:= \mathsf{T} \\ \mathsf{S} n =_\mathbb{N} 0 &:= \mathsf{F} \\ 0 =_\mathbb{N} \mathsf{S} m &:= \mathsf{F} \\ \mathsf{S} n =_\mathbb{N} \mathsf{S} m &:= n =_\mathbb{N} m \end{aligned}$$

Equality for booleans can also be easily defined since it is just a boolean operator.

We say two terms are *convertible* if they reduce to the same term. We write $t \leftrightsquigarrow s$ to mean that the two terms $t$ and $s$ are convertible. Convertible terms denote the same value, so if $t \leftrightsquigarrow s$ then we can always replace $t$ with $s$ in any context.

For the logic, we will assume we are given symbols $\top$, $\bot$, $\wedge$, $\Rightarrow$[2], and $\forall$. We will take as inference rules for these symbols the standard rules given by natural deduction [Troelstra and Schwichtenberg, 1996]. When a formula is derivable using the rules of our logic, we will say that the formula is *valid*, or the formula *holds*. This collection of classical connectives is well known to be complete. We can define the remaining logical connectives in terms of these:

$$\begin{aligned} \neg \varphi &:= \varphi \Rightarrow \bot \\ \varphi \vee \psi &:= \neg(\neg \varphi \wedge \neg \psi) \\ \exists a\colon A.\varphi(a) &:= \neg \forall a\colon A. \neg \varphi(a) \end{aligned}$$

We will call $\vee$ the *classical disjunction* or *weak disjunction*, and we will call $\exists$ the *classic existential* quantifier or *weak existential* quantifier.

For atomic relations, we define only one predicate symbol over the booleans, $\langle \cdot \rangle$:

$$\begin{aligned} \langle \mathsf{F} \rangle &:= \bot \\ \langle \mathsf{T} \rangle &:= \top \end{aligned}$$

Notice that we are distinguishing between the type $\mathbb{B}$ of two elements and logical propositions. The $\langle \cdot \rangle$ relation acts as a kind of conversion from booleans to propositions.

We will not take any primitive notion of equality. Instead, we will use $\langle r =_\mathbb{N} s \rangle$ for the equality relation between two natural numbers and $\langle r =_\mathbb{B} s \rangle$ for the equality relation between two booleans.

We will assume we are given rules for case analysis and induction:

$$\frac{\theta(\mathsf{T}) \quad \theta(\mathsf{F})}{\forall b\colon \mathbb{B}.\theta(b)} \qquad\qquad \frac{\theta(0) \quad \begin{array}{c}[n\colon \mathbb{N};\ \theta(n)] \\ \vdots \\ \theta(\mathsf{S} n)\end{array}}{\forall n\colon \mathbb{N}.\theta(n)}$$

**Figure 1.** Rule for case analysis.      **Figure 2.** Rule for induction.

To see that we have classical logic, we prove the law of excluded middle.

---

1. Formally, equality is defined as

$$\lambda n.\operatorname{rec}_{\mathbb{N} \Rightarrow \mathbb{B}} n\, (\lambda m.\operatorname{rec}_\mathbb{B} m\, \mathsf{T}\, (\lambda m_0 c.\mathsf{F}))\, (\lambda n_0\, r m.(\operatorname{rec}_\mathbb{B} m\, \mathsf{F}\, (\lambda m_0 c. r m_0))).$$

2. Context can be used to disambiguate whether $\Rightarrow$ refers to implication or a function type. We use the same symbol because in Section 3 we will actually identify implication with the function type.

**Theorem 1.** *(Law of excluded middle) For any proposition $\varphi$, $\varphi \vee \neg\varphi$ holds.*

**Proof.** By the definition of $\vee$, we need to prove $\bot$ from the assumptions of $\neg\varphi$ and $\neg\neg\varphi$. But $\neg\neg\varphi = \neg\varphi \Rightarrow \bot$. By applying modus ponens to $\neg\varphi$ and $\neg\varphi \Rightarrow \bot$, we get $\bot$ as required. $\square$

There is an objection that can be raised here. Even though we have proved the law of excluded middle, we still do not have a theorem giving us proof by contradiction. More specifically, there is no proof schema for $\neg\neg\varphi \Rightarrow \varphi$ in our logic. This is not a problem because we can prove this for every formula in our classical fragment, specifically those formulas built from, $\top$, $\bot$, $\langle \cdot \rangle$, $\wedge$, $\Rightarrow$, and $\forall$, and hence the other connectives that are defined in terms of these.

**Theorem 2.** *(Proof by contradiction) For any propositions $\varphi$ and $\psi$ and for any term $r$, the following hold.*

1. $\neg\neg\top \Rightarrow \top$
2. $\neg\neg\bot \Rightarrow \bot$
3. $(\neg\neg\varphi \Rightarrow \varphi) \Rightarrow (\neg\neg\psi \Rightarrow \psi) \Rightarrow (\neg\neg(\varphi \wedge \psi) \Rightarrow (\varphi \wedge \psi))$
4. $(\neg\neg\psi \Rightarrow \psi) \Rightarrow (\neg\neg(\varphi \Rightarrow \psi) \Rightarrow (\varphi \Rightarrow \psi))$
5. $\neg\neg\langle r \rangle \Rightarrow \langle r \rangle$
6. $(\forall a \colon A. \neg\neg\varphi(a) \Rightarrow \varphi(a)) \Rightarrow (\neg\neg(\forall a \colon A.\varphi(a)) \Rightarrow \forall a \colon A.\varphi(a))$

**Proof.** (1)–(4) are easy constructive tautologies. By case analysis (5) reduces to cases (1) and (2). For (6) we need to prove $\varphi(a)$ for an arbitrary $a$ given our hypothesis. From $\forall a.\neg\neg\varphi(a) \Rightarrow \varphi(a)$ it suffices to prove $\neg\neg\varphi(a)$, but this follows easily from $\neg\neg(\forall a.\varphi(a))$. $\square$

We say a proposition $\varphi$ is *double negation stable*, or simply *stable* if $\neg\neg\varphi \Rightarrow \varphi$ holds. If $\varphi$ is a stable formula then we are allowed to use proof by contradiction to prove it. By induction on the syntax of formulas, we see that every formula we can build so far is stable, and hence proof by contradiction is available to prove it. Notice that for $\varphi \Rightarrow \psi$ to be stable we only require that $\psi$ be stable. Therefore, hypotheses of a proposition are irrelevant when determining whether it is stable or not; only the conclusion matters.

With this fragment of our logic so far, we have covered all of higher-order Peano arithmetic. This means that this fragment has more than enough logical power to develop essentially all of classical number theory using classical reasoning and even most of analysis [Simpson, 1999]. This way of defining classical logic inside constructive logic is known as the Gödel-Gentzen double negation interpretation [Gödel, 1933].

Now we turn our attention to extending this logic with constructive connectives.

## 2.2 Constructive logic

We will add two new primitive symbols: the constructive disjunction $+$[3], and the constructive existential quantifier $\Sigma$. These connectives come with their standard rules of inference provided by natural deduction.

With the introduction of these two connectives, our inductive proof that $\neg\neg\varphi \Rightarrow \varphi$ now breaks down. We cannot expect $\neg\neg(\varphi + \psi) \Rightarrow (\varphi + \psi)$ to hold in general even when $\varphi$ and $\psi$ are stable. Neither can we prove in general that $\neg\neg(\Sigma n \colon \mathbb{N}. \varphi(n)) \Rightarrow (\Sigma n \colon \mathbb{N}.\varphi(n))$ holds even when $\varphi(n)$ is stable for all $n \colon \mathbb{N}$.

We say a proposition $\varphi$ is *decidable* if $\varphi + \neg\varphi$ holds. Every decidable proposition is stable.

Recall equations (1) and (2) which state that our new constructive connectives can be weakened to their classical counterparts. This reflects part of the old view of constructive logic where every constructive proof is a classical proof. However, this old view is not fully internalized in constructive logic. We do not have that every proposition with constructive connectives implies the same proposition with the constructive connectives replaced by their classical counterparts.

---

[3]. Context will be use to disambiguate the constructive disjunction from addition.

**Theorem 3.** *The formula $\neg(\forall n\colon \mathbb{N}.\Sigma m\colon \mathbb{N}.R n m) \Rightarrow \neg(\forall n\colon \mathbb{N}.\exists m\colon \mathbb{N}.R n m)$ does not hold in general.*

**Proof.** Given our deduction rules, we know that every valid formula holds in every topological model [Rasiowa. and Sikorski, 1968]. Therefore, to prove that this formula is not valid, it suffices to find a topological model where the formula does not hold.

In the topological model we will consider, propositions will represent open subsets of $\mathcal{C}$, the Cantor space. Let $f\colon \mathbb{N} \Rightarrow \mathcal{C}$ be an enumeration of some countable dense subset of the Cantor space. Let $R n m$ be clopen sets such that $\bigcup_{m \in \mathbb{N}} R n m$ covers the entire Cantor space except $f(n)$. Thus $\Sigma m.R n m$ represents $\mathcal{C} \setminus \{f(n)\}$, but $\exists m.R n m$ represents $\mathcal{C}$. Therefore $\forall n.\Sigma m.R n m$ represents $\emptyset$ while $\forall n.\exists m.R n m$ represents $\mathcal{C}$. Thus $\neg(\forall n.\Sigma m.R n m)$ represents $\mathcal{C}$ and $\neg(\forall n.\exists m.R n m)$ represents $\emptyset$. Hence the theorem is false in this model. □

Even adding an assumption that $R$ is decidable does not help.

**Theorem 4.** *The formula*

$$(\forall n m\colon \mathbb{N}.R n m + \neg R n m) \Rightarrow \neg(\forall n\colon \mathbb{N}.\Sigma m\colon \mathbb{N}.R n m) \Rightarrow \neg(\forall n\colon \mathbb{N}.\exists m\colon \mathbb{N}.R n m)$$

*does not hold constructively.*

**Proof.** The topological model from Theorem 3 satisfies the hypothesis $\forall n m\colon \mathbb{N}.R n m + \neg R n m$, so the proof is the same as the proof of Theorem 3. □

## 2.3 Reasoning with weak connectives

One wants to reason with the weak connectives as easily as one reasons with the connectives in traditional classical logic. Unfolding the definition of the weak disjunction to transform it into a negated conjunction is not a pleasant way of proceeding. One wants to do case analysis. We cannot allow general case analysis to be done on the weak disjunction because that would imply that it is equivalent to the constructive disjunction.

Fortunately, there is a common situation when case analysis is allowed on the weak disjunction. When the goal is a stable proposition, then case analysis can be done:

$$\cfrac{\varphi \vee \psi \quad \begin{array}{c}[\varphi]\\ \vdots\\ \theta\end{array} \quad \begin{array}{c}[\psi]\\ \vdots\\ \theta\end{array} \quad \begin{array}{c}[\neg\neg\theta]\\ \vdots\\ \theta\end{array}}{\theta}$$

**Figure 3.** A derived elimination rule for weak disjunction and stable $\theta$.

It is easy to prove this rule is a tautology in constructive logic. A similar rule can be derived for the weak existential:

$$\cfrac{\exists a\colon A.\varphi(a) \quad \begin{array}{c}[a\colon A;\, \varphi(a)]\\ \vdots\\ \theta\end{array} \quad \begin{array}{c}[\neg\neg\theta]\\ \vdots\\ \theta\end{array}}{\theta}$$

**Figure 4.** A derived elimination rule for the weak existential and stable $\theta$.

In Coq, these two rules, when written as theorems, have a form close enough to that of an elimination rule of an inductive type that they can be used with Coq's `destruct t using` tactic. This allows case analysis to be done on weak disjunctions and weak existentials almost as easily as they are done on constructive disjunctions and constructive existentials. To make this easier still, Appendix A provides tactics (`orWelim` and `existWelim`) that calls `destruct t using` and then tries to automatically solve the stability goal for $\theta$ by searching a database of stability hints.

# 3 Dependent Type Theory

Up to now, we have been taking a traditional view of logic where terms and formula are separate entities. However, to proceed will will need to take a unified view of these two entities, and that unified view is provided by dependent type theory.

In dependent type theory, both data types and logical propositions are put on the same level; both are considered types. We define $\star$ as a kind of types[4] and we can write $\mathbb{N}\colon \star$, or $\forall n\colon \mathbb{N}.\Sigma m\colon \mathbb{N}.R\,n\,m\colon \star$. The values of propositional types are *proof objects*, or witnesses, of the proposition. For example, $\lambda x\colon \varphi.x$ has type $\varphi \Rightarrow \varphi$, and it is a proof object of the this tautology. For every deduction of a formula $\theta$, there is an associated proof object of type $\theta$ which is given by the Curry-Howard isomorphism.

Dependent type theory enhances the elimination rules for logical connectives allowing dependent elimination. For example, the dependent elimination rule for the constructive existential is shown in Figure 5.

$$\frac{p\colon \Sigma a\colon A.\varphi(a) \quad \begin{array}{c}[a\colon A;\, \varphi(a)]\\ \vdots\\ \theta((a, \varphi(a))_{\Sigma\varphi})\end{array}}{\theta(p)}$$

**Figure 5.** Dependent elimination rule for the constructive existential.

Here the proof object $p$ is allowed to occur in the type of the conclusion, $\theta(p)$. Notice that when applying the elimination rule to a goal $\theta(p)$, the $p$ occurring in $\theta$ is refined into the dependent pair $(a, \varphi(a))_{\Sigma\varphi}$. This is analogous to the induction rule for natural numbers where the $n$ occurring in $\theta(n)$ is refined into $0$ and $\mathsf{S}m$ in the two branches. In fact, it is completely analogous, so much so that we can define all the logical connectives to actually be data types.

$$\begin{array}{rcll} \top & := & () & \text{(unit type)}\\ \bot & := & \emptyset & \text{(void type)}\\ \varphi \wedge \psi & := & \varphi \times \psi & \text{(product type)}\\ \varphi \Rightarrow \psi & := & \varphi \Rightarrow \psi & \text{(function type)}\\ \forall x\colon A.\varphi(a) & := & \Pi x\colon A.\varphi(a) & \text{(dependent function type)} \end{array}$$

The classical formulas correspond to "degenerate" data types, in the sense that if $\varphi$ is a proposition made entirely from classical connectives,[5] then for any two values of such a type, $p, q\colon \varphi$, we can prove $p \asymp q$ for a suitable notion of equivalence (see Section 3.1). Generally, we will use the logical notation for these classical formulas, and reserve the use of the type theoretic notation for the other cases. However, technically they will denote the same thing.

What we have called the constructive disjunction and constructive existential are really the same things as the sum type and dependent pair type respectively, and we will continue to use the same notation for them both.

Since we have a kind for types, we can give signatures to predicates and relations. A predicate over $A$ will have kind $A \Rightarrow \star$, and a binary relation on $A$ and $B$ will have kind $A \Rightarrow B \Rightarrow \star$, which is the same thing as a function from $A$ to predicates on $B$. We already have an example of a predicate: $\lambda b\colon \mathbb{B}.\langle b \rangle \colon \mathbb{B} \Rightarrow \star$. For $b\colon \mathbb{B}$, we can now define $\langle b \rangle$ as

$$\langle b \rangle := \text{if } b \top \bot.$$

This basic predicate can be used build other, more complicated predicates.

---

4. There are many variations of dependent type theory which treat the types of types in various ways, and there are choices about whether quantification is predicative or impredicative. In this paper, we shall ignore these issues and leave it up to the reader to interpret this work as much as possible in whatever system they wish. For concreteness, this work has been prototyped in Coq. See Appendix A.

5. Actually, it suffices that the conclusion be composed from classical connectives.

In exactly the same way we can define families of types. For example, we can recursively define a family of types having $n\colon \mathbb{N}$ elements as a function of $n$:

$$\begin{aligned} \mathbb{Z}_0 &:= \emptyset \\ \mathbb{Z}_{\mathsf{S}n} &:= () + \mathbb{Z}_n. \end{aligned}$$

## 3.1 Setoids

So far we have not dealt with equality beyond the recursive definition of equality for inductive data types such as $\mathbb{N}$. In dependent type theory, it is common to use a structure called a *setoid* that pairs a data type, called the carrier, with an equivalence relation over that type. The type of Setoids is denoted by $\Omega$.

$$\Omega := \Sigma A\colon \star.\, \Sigma R\colon A \Rightarrow A \Rightarrow \star.\, \mathrm{EquivalenceRelation}\, R$$

When a $X\colon \Omega$ is used in a context expecting a type, we will leave the first projection implicit. We will write $x \asymp_X y$ when $x$ and $y$ are equivalent under the equivalence relation from the setoid $X$. We will leave the subscript off when it can be inferred from context.

All inductive data types form a natural setoid structure where equality is the one defined in the usual recursive way. These natural setoid equivalences for the void data type $\emptyset$ and the unit data type $()$ are trivial in the sense that equivalences hold for every pair of values in their respective types.

A function $f\colon X \Rightarrow Y$ between two setoids $X$ and $Y$ is respectful when it respects the equivalence relation from the two setoids:

$$\mathrm{Respectful}\, f := \forall x\, y\colon X.\, x \asymp y \Rightarrow f x \asymp f y$$

The type of all respectful functions forms the (constructive) function space between setoids $X$ and $Y$. We write $X \to Y$ for this type:

$$X \to Y := \Sigma f\colon X \Rightarrow Y.\, \mathrm{Respectful}\, f$$

In fact, we will make $X \to Y$ a setoid. We define two respectful functions $f,\, g\colon X \to Y$ to be equivalent if they are equivalent pointwise:

$$f \asymp g := \forall x\colon X.\, f x \asymp g x.$$

Propositions also form a setoid. Two propositions $\varphi,\, \psi\colon \star$ are considered equivalent if they are logically equivalent:

$$\varphi \asymp \psi := (\varphi \Rightarrow \psi) \wedge (\psi \Rightarrow \varphi)$$

With this together with respectful functions, we can form respectful predicates of type $X \to \star$ and higher arity respectful relations.

Given a setoid $X$ and a respectful predicate on that setoid, $P\colon X \to \star$, we can define a subsetoid with carrier $\Sigma a\colon X.\, P a$ and where equality is defined by equality of $X$ on the first projection.

$$(x, p) \asymp_{\Sigma P} (y, q) := x \asymp_X y$$

Given a setoid $X$ and a respectful equivalence relation on that setoid $E$, a quotient setoid can be formed by replacing the equivalence relation on $X$ with $E$.

$$x \asymp_{X/E} y := E\, x\, y$$

In this paper, we will also be interested in *stable setoid*s, which are setoids that have an additional property that the equivalence relation is stable.[6] Setoids for inductive data types are stable setoids, and respectful functions whose codomain are stable setoids are also stable setoids. One can define a setoid of (constructive) real numbers [O'Connor, 2008a], $\mathbb{R}$, and this setoid is stable.

---

6. Stable setoids could reasonably be called Hausdorff spaces [Bauer and Taylor, 2008].

## 3.2 Notation

We will use some compressed notation for a few common idioms. In particular, we will identify sets with respectful predicates, so we will use set notation to define and work with predicates:

$$\begin{aligned}
\wp X &:= X \to \star \\
x \in X &:= X x \\
\{x\colon X \,|\, \varphi(x)\} &:= \lambda x\colon X.\,\varphi(x) \\
\{fx \,|\, x\colon X\} &:= \{y \,|\, \Sigma x\colon X.\,y \asymp fx\} \\
X \cup Y &:= \{x \,|\, x \in X + x \in Y\} \\
X \cap Y &:= \{x \,|\, x \in X \land x \in Y\} \\
X \subseteq Y &:= \forall x.\,x \in X \Rightarrow x \in Y
\end{aligned}$$

We also define some shorthand for quantifiers:

$$\begin{aligned}
\forall x \in X.\,\varphi(x) &:= \forall x.\,x \in X \Rightarrow \varphi(x) \\
\Sigma x \in X.\,\varphi(x) &:= \Sigma x.\,x \in X \land \varphi(x) \\
\forall n < m.\,\varphi(n) &:= \forall n.\,\langle n < m \rangle \Rightarrow \varphi(n) \\
\Sigma n < m.\,\varphi(n) &:= \Sigma n.\,\langle n < m \rangle \land \varphi(n)
\end{aligned}$$

and similarly for $\exists$ and for other inequalities.

Finally, we define unique existence for setoids.

$$\exists! x\colon X.\,\varphi(x) := (\exists x\colon X.\,\varphi(x)) \land (\forall x\,y\colon X.\,\varphi(x) \Rightarrow \varphi(y) \Rightarrow x \asymp y)$$

In most cases we will leave the first projection of the dependent pair type (a.k.a. the constructive existential) implicit as I have already mentioned in the case of using setoids in a context requiring a type.

I will also not explicitly write out proof objects that go into the construction of constructive existentials. Instead, I will just give the witness and will leave it to the reader to fill in the proof.

# 4 Infinite Pigeonhole Principle

Let us consider a real life example of using classical reasoning in the development of constructive mathematics. In this example, we will be using the infinite pigeonhole principle. The infinite pigeonhole principle says that if you distribute an infinite number of pigeons amongst a finite number of pigeonholes, then there is some pigeonhole with an infinite number of pigeons. There is no constructive proof of the infinite pigeonhole principle because finding a pigeonhole with an infinite number of pigeons is undecidable in general. Therefore, we state the theorem using classical existentials.

$$\forall n\colon \mathbb{N}.\,\forall f\colon \mathbb{N} \Rightarrow \mathbb{Z}_n.\,\exists m.\,\forall a\colon \mathbb{N}.\,\exists b > a.\,\langle fb =_{\mathbb{Z}_n} m \rangle.$$

Because this theorem is stated in the classical fragment of our logic, we can use classical reasoning to prove this theorem.

Now let us see a constructive context where this classical theorem can be used.

## 4.1 Hausdorff metrics

I have used this pigeonhole principle during my development of constructively compact sets to prove that the Hausdorff metric is, in fact, a metric [O'Connor, 2008b, O'Connor, 2009]. A metric on a type $X$ is defined by a ball relationship $B^X\colon \mathbb{Q}^+ \Rightarrow X \Rightarrow X \Rightarrow \star$ where $B_\varepsilon^X a b$ means that $d\,a\,b \leqslant \varepsilon$. The ball relationship is used instead of the distance function for a couple of reasons. Firstly, this allows one to define a metric without needing to construct the real numbers. Secondly, it allows one to define metrics for spaces without a computable distance function.

Given a metric space $X$, one can define a metric on $\mathfrak{F}X$, the finite sets of $X$, by the Hausdorff metric. Given two finite sets $\mathcal{A}$ and $\mathcal{B}$, we defined the ball relation as

$$B_\varepsilon^{\mathfrak{F}X} \mathcal{A}\mathcal{B} := (\forall a \in \mathcal{A}.\,\exists b \in \mathcal{B}.\,B_\varepsilon^X a b) \land (\forall b \in \mathcal{B}.\,\exists a \in \mathcal{A}.\,B_\varepsilon^X a b).$$

This says that for every point in $\mathcal{A}$ there is some point in $\mathcal{B}$ that within $\varepsilon$ and vice versa. Notice the use of the weak existential here. It may be unclear why one needs use the weak existential and not the constructive one. To understand, it is helpful to consider the case where $X$ is the real numbers $\mathbb{R}$.

If we were using the traditional metric space definition, the distance between two sets of real numbers $\mathcal{A}$ and $\mathcal{B}$ would be

$$d\mathcal{A}\mathcal{B} := \max\left\{\max_{a \in \mathcal{A}} \min_{b \in \mathcal{B}} d\,a\,b, \max_{b \in \mathcal{B}} \min_{a \in \mathcal{A}} d\,a\,b\right\}.$$

Since $\mathcal{A}$ and $\mathcal{B}$ are finite sets and min and max are constructive functions, this definition is constructive. However, notice that $\min_{b \in \mathcal{B}} d\,a\,b$ does not compute at which point $b$ the minimum value is obtained. If such a $b$ were computable then we could decide for any two real numbers if $a \leqslant b$ or $b \leqslant a$. However, this is well known to be undecidable.

Since we want to represent this metric using our ball relation, it would be inappropriate to use constructive existence in our Hausdorff metric definition since that would require us to know at which point $b$ the minimum $\min_{b \in \mathcal{B}} d\,a\,b$ is obtained. This would make the definition too strong.

## 4.2 Applying the pigeonhole principle

One key property required of the ball relation is the closedness property:

$$\forall \varepsilon\colon \mathbb{Q}^+.(\forall \delta\colon \mathbb{Q}^+. B^X_{\varepsilon+\delta}ab) \Rightarrow B^X_\varepsilon ab$$

We need to prove this property for our definition of the Hausdorff metric (given that the metric space $X$ already has this property). Consider for the moment one half of the definition of the Hausdorff metric:

$$\forall a \in \mathcal{A}. \exists b \in \mathcal{B}. B^X_\varepsilon ab$$

To prove half of the closeness property, it suffices to prove

$$\forall \varepsilon\colon \mathbb{Q}^+.\left(\forall n\colon \mathbb{N}^+.\forall a \in \mathcal{A}.\exists b \in \mathcal{B}. B^X_{\varepsilon+\frac{1}{n}}ab\right) \Rightarrow \forall a \in \mathcal{A}.\exists b \in \mathcal{B}. B^X_\varepsilon ab.$$

For any given $a \in \mathcal{A}$ we need to prove there is some $b \in \mathcal{B}$ such that $B^X_\varepsilon ab$ holds. The trouble is that the $b$ found by our hypothesis depends on $n$, and we need to find one $b$ that holds for every $n$.

This is where the infinite pigeonhole principle comes into play. There is going to be some $b$ such that that $B^X_{\varepsilon+\frac{1}{n}}ab$ holds for infinitely many $n$ since $\mathcal{B}$ is a finite set. Because of the weakening property of metric spaces, this one $b$ works for all $n$. Thus $B^X_\varepsilon ab$ holds as needed. The other half of the definition of the Hausdorff metric is proved similarly. Notice that because we have used the weak existential in the definition of the Hausdorff metric, we are able to apply the infinite pigeonhole principle, which is a result in classical logic.

One might object to this whole formulation of the problem because, though it is technically done in constructive logic, we have avoided any real constructive work by avoiding the constructive existential. In other words, there is no computational content in any of this work.

While not denying the above argument, it is important to see that we still maintain the constructive content of the rest of the theory. While there is no computational content in this fragment of the development, once the Hausdorff metric is defined one can go on to define compact sets as the Cauchy completion of the finite sets.

The completion of a metric space $X$ [O'Connor, 2008a] is represented by coherent functions from $\mathbb{Q}^+$ to $X$. A function representing a point $y$ from the completion of $X$ takes a tolerance $\varepsilon\colon \mathbb{Q}^+$ and returns a value in $X$ that is within $\varepsilon$ of the point $y$. A function is *coherent* if it represents some point. Two different functions that both represent the same point are equivalent under a suitable setoid equivalence relation. We see that the completion does have constructive content, even if the underlying metric is in the classical fragment of constructive logic. For example, for each compact subset of the plane, we can compute a finite set of points that approximates the compact set.

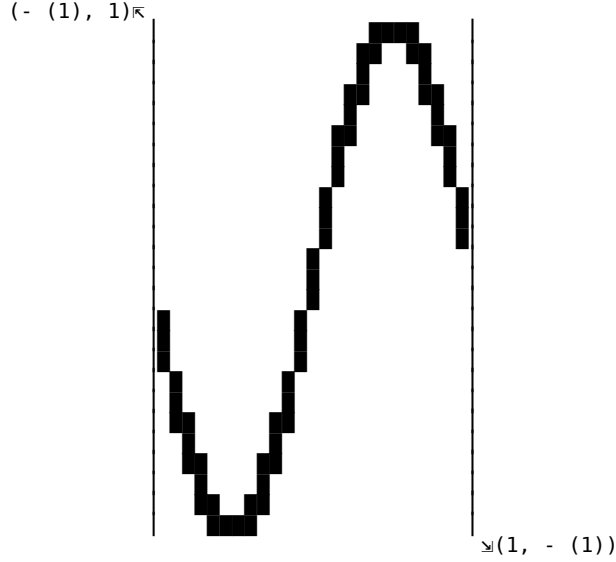

**Figure 6.** The result of computing an approximation of the graph of $\sin(3x)$ on $[-1,1]$.

Figure 6 shows the output of a computation inside Coq (not extracted from Coq) of an approximation of the graph of the function $\lambda x.\sin(3x)$ on the interval $[-1,1]$. Here the finite set is being represented as a boolean matrix together with coordinates for the upper-left and lower-right corners of the matrix. Using the notation mechanism of Coq we display the boolean matrix using single Unicode characters denoting whether the value of the corresponding entry of the matrix is `true` or `false`. This produces the convenient visualization of the finite set seen in the figure and illustrates how concrete computation can really proceed even when the metric is in the classical fragment of our logic.

## 5 Double Negation

There is an alternative definition of the classical disjunction and classical existential in terms of their constructive counter parts. By slapping a double negation on the outside of a constructive connective, it can be turned into a classical connective.

$$\varphi \vee \psi := \neg\neg(\varphi + \psi)$$
$$\exists x.\varphi(x) := \neg\neg(\Sigma x.\varphi(x))$$

These definitions are equivalent to the definitions given in previously in Section 2.1, but the observation that you use double negations to transform constructive statements into classical ones is very powerful owing to the special properties of double negation. In particular, double negation forms a monad, and thus monadic style reasoning can be used.

### 5.1 Monads

Monads are a standard construction in category theory [Kock, 1972]. In this section, we review the basics of monads from a functional programming perspective [Wadler, 1995]. A (strong) monad $\mathfrak{M}\colon \Omega \Rightarrow \Omega$ is a setoid constructor that comes with three polymorphic operations.

$$\begin{aligned}
\text{unit}_{\mathfrak{M}} &: \Pi X. X \to \mathfrak{M} X \\
\text{map}_{\mathfrak{M}} &: \Pi XY.(X \to Y) \to \mathfrak{M} X \to \mathfrak{M} Y \\
\text{join}_{\mathfrak{M}} &: \Pi X. \mathfrak{M}(\mathfrak{M} X) \to \mathfrak{M} X
\end{aligned}$$

These operations need to satisfy certain laws. There are several different ways of phrasing the monad laws, but one of the easiest ways is to first define two auxiliary combinators. The first combinator, $\text{bind}_{\mathfrak{M}}$ takes a function $f\colon X \to \mathfrak{M} Y$ and lifts the argument, returning a function of type $\mathfrak{M} X \to \mathfrak{M} Y$:

$$\text{bind}_{\mathfrak{M}} f := \text{join}_{\mathfrak{M}} \circ (\text{map}_{\mathfrak{M}} f)$$

The second combinator takes $f\colon X \to \mathfrak{M}Y$ and $g\colon Y \to \mathfrak{M}Z$ and combines them to get a function of type $X \to \mathfrak{M}Z$:

$$g \lll f := (\text{bind}_{\mathfrak{M}}\, g) \circ f$$

Using this combinator, the monad laws can be stated as

$$\begin{aligned} f \lll \text{unit}_{\mathfrak{M}} &\asymp f \\ \text{unit}_{\mathfrak{M}} \lll f &\asymp f \\ f \lll (g \lll h) &\asymp (f \lll g) \lll h. \end{aligned}$$

Let us call a function whose type is of the form $X \to \mathfrak{M}Y$ a Kleisli function. As you can see, these Kleisli functions compose together nicely. By composing a regular function of type $X \to Y$ together with $\text{unit}_{\mathfrak{M}}$, it can be transformed into Kleisli function.

We will use the following notation for these operations when it is clear which monad is under consideration:

$$\begin{aligned} \bar{f} &:= \text{map}_{\mathfrak{M}}(f) \\ \check{f} &:= \text{bind}_{\mathfrak{M}}(f) \\ \hat{a} &:= \text{unit}_{\mathfrak{M}}(a) \end{aligned}$$

There are several other useful combinators for monads. One called ap, takes a function "inside a monad", $f\colon \mathfrak{M}(X \to Y)$, and a value $x\colon \mathfrak{M}X$ "applies" $f$ to $x$ and returns a value of type $\mathfrak{M}Y$. We write ap using the infix operator @:

$$f \mathbin{@} x := \text{bind}_{\mathfrak{M}}\, (\lambda f_0.\, \bar{f}_0\, x)\, f$$

Ap can be used to create higher arity map functions, such as

$$\text{map2}_{\mathfrak{M}}\colon (X \to Y \to Z) \to \mathfrak{M}X \to \mathfrak{M}Y \to \mathfrak{M}Z$$

$$\text{map2}_{\mathfrak{M}}\, f\, x\, y := \hat{f} \mathbin{@} x \mathbin{@} y.$$

We will give $\text{map2}_{\mathfrak{M}}$ a special notation when applied to infix operators:

$$x \,\bar{\diamond}\, y := \text{map2}_{\mathfrak{M}}\, (\lambda a\, b.\, a \diamond b)\, x\, y$$

With these combinators, plus other combinators and notations, manipulating monadic values can be almost as pleasant as manipulating regular values.

## 5.2 The double negation monad

Double negation $\mathfrak{N}\colon \star \Rightarrow \star$ is a type constructor:

$$\mathfrak{N}\varphi := \neg\neg\varphi$$

This is trivially a setoid constructor because every function to $\emptyset$[7] is respectful and equivalent to every other function of the same type.

The operations for this monad correspond to to the natural proofs of the appropriate type:

$$\begin{aligned} \text{unit}_{\mathfrak{N}} &:= \lambda \varphi\colon \star.\, \lambda a\colon \varphi.\, \lambda k\colon \neg\varphi.\, k\, a \\ \text{map}_{\mathfrak{N}} &:= \lambda \varphi\, \psi\colon \star.\, \lambda f\colon \varphi \Rightarrow \psi.\, \lambda x\colon \mathfrak{N}\varphi.\, \lambda k\colon \neg\psi.\, x\, (k \circ f) \\ \text{join}_{\mathfrak{N}} &:= \lambda \varphi\colon \star.\, \lambda x\colon \mathfrak{N}(\mathfrak{N}\varphi).\, \lambda k\colon \neg\varphi.\, x\, (\text{unit}_{\mathfrak{N}}\, k) \end{aligned}$$

Any double negated type, $\mathfrak{N}\varphi$, is automatically stable no matter if $\varphi$ uses constructive connectives or not. Indeed, $\text{join}_{\mathfrak{N}}$ witnesses the proof of stability. This means that weak disjunction and weak existentials can be eliminated nicely whenever one is trying to prove a goal of type $\mathfrak{N}\theta$. This is actually a special instance of a more general rule that says that if the goal stable then a hypothesis of the form $\mathfrak{N}\varphi$ can be transformed into a hypothesis of the form $\varphi$.

---

7. Recall that $\neg\varphi := \varphi \Rightarrow \bot$ and that $\bot := \emptyset$, so a value of type $\neg\varphi$ is really a function to the void type.

$$\frac{\mathfrak{N}\varphi \quad \overset{[\varphi]}{\underset{\theta}{\vdots}} \quad \overset{[\neg\neg\theta]}{\underset{\theta}{\vdots}}}{\theta}$$

**Figure 7.** Elimination rule for the double negation monad and stable $\theta$.

In the special case where $\theta$ is of the form $\mathfrak{N}\theta_0$, the proof of this elimination rule is witnessed by the $\text{bind}_\mathfrak{N}$ function. Indeed, if $\theta$ is stable, then $\theta$ and $\mathfrak{N}\theta$ are isomorphic, with $\text{unit}_\mathfrak{N}$ witnessing $\theta \Rightarrow \mathfrak{N}\theta$, and the proof of stability witnessing $\mathfrak{N}\theta \Rightarrow \theta$. Using this isomorphism, we can prove the our elimination rule for the double negation monad by transforming $\theta$ into $\mathfrak{N}\theta$, using $\text{bind}_\mathfrak{N}$, and transforming $\mathfrak{N}\theta$ back to $\theta$.

The advantage of this monadic approach is that, not only can classical connectives for disjunction and existence be defined, but arbitrarily complex constructive propositions can be turned into classical-like propositions.

## 5.3 Hybrid systems

Geuvers et al. have developed a framework for proving safety of hybrid discrete and continuous systems [Geuvers etal., 2010]. For example, they prove the correctness of a simple thermostat hybrid system by proving that it maintains a minimum temperature.

This work uses exact real arithmetic via the constructive real numbers, rather than floating point numbers, in order to ensure correctness. The difficulty here is that many propositions about real numbers are undecidable and hence not provable in constructive mathematics. For example, constructive trichotomy $\forall x\,y\colon \mathbb{R}.\,(x < y) + (x \asymp y) + (y < x)$, and even constructive dichotomy $\forall x\,y\colon \mathbb{R}.\,(x \leqslant y) + (y \leqslant x)$ do not hold, but they do hold when the constructive disjunction is replaced by the weak disjunction or when the lemma conclusions are put into the double negation monad.

Geuvers et al. use the double negation monad instead of classical connectives. The use of the double negation monad allows one to use trichotomy and dichotomy when reasoning about the hybrid systems. Using the classical fragment of constructive logic is acceptable because the ultimate goal of a safety proof is to show that the set of reachable states is contained within some compact set, and being a member of a compact set is a stable relation.

## 5.4 Feit-Thompson theorem

The Feit-Thompson theorem, also known as the odd order theorem, states that any group of odd order is solvable. The revised paper proof is about 255 pages long and there is currently an ambitious effort underway to formalize the proof in Coq [Gonthier etal., 2007].

The ultimate goal of the Feit-Thompson theorem is double negation stable, thus classical reasoning can be allowed. Since large parts of the proof only involve reasoning over finite sets, restricting oneself to constructive reasoning is not actually restrictive because most classical laws hold constructively in the finite case. However, a few parts of the proof involve reasoning about undecidable predicates. For these cases, a modal operator called `classically` has been defined:[8]

$$\texttt{classically}\,\varphi := \forall b\colon \mathbb{B}.\,(\varphi \Rightarrow \langle b \rangle) \Rightarrow \langle b \rangle$$

Notice that this definition is logically equivalent to the double negation monad defined in Section 5.2. The reason this form of the double negation monad is used is that, for this particular project, one very often has goals of the form $\langle r \rangle$. When this is the case a hypothesis of `classically` $\varphi$ can be immediately applied and a new hypothesis $\varphi$ introduced into the context. Using SSReflect tactics [Gonthier and Stéphane, 2009], this can easily be done in one step. This leaves the user to prove their original goal $\langle r \rangle$ under the hypothesis $\varphi$ with the `classically` stripped away.

---

8. Technically the definition of `classically` is a little different than the one we give here, but the spirit is close enough for our purposes.

# 6 Classical Functions

In classical mathematics, functions are represented by a certain class of binary predicates over the domain and codomain. The constrains are that for each point in the domain there is at at least one and at most one related point in the codomain.

$$(\forall x\colon X. \exists Y. Fxy) \land (\forall yz\colon B. Fxy \Rightarrow Fxz \Rightarrow y \asymp z).$$

If we drop the constraint that there is at least one related point in the codomain we get the definition of a partial function.

Defining this notion of a classical function using constructive mathematics is not difficult. The problem is that we want to be able to compose and manipulate functions easily, and using relations makes this difficult. For example, to write $F(Gx) \asymp G(Fx)$ using binary predicates, one would have to give names to the intermediate values and write out something like

$$\exists y_1\, y_2\, z_1 z_2. Gxy_1 \land Fy_1\, z_1 \land Fxy_2 \land Gy_2\, z_2 \land z_1 \asymp z_2.$$

Translating from functional notation to binary predicate notation is quite painful and greatly expands the formula. In first order logic, there is not much else that can be done without extending the language with new function symbols. However, with dependent type theory there is an easy transformation that can be done.

Using curried functions, one would write the type of a binary predicate on $X$ and $Y$ as $X \to Y \to \star$. Binary predicates can also be equivalently viewed as having type $X \to \wp Y$, which is a function from $X$ to unary predicates on $Y$.

If the binary predicate $F$ satisfies the properties of being a function and it is applied to an input $x\colon X$, the resulting unary predicate will have some special properties. The predicate $Fx\colon \wp Y$ is a predicate that holds for at least one value and at most one value. For a classical partial function $F$, the predicate $Fx$ will hold for at most one value.

We define a *classical partial value*, or a *weak partial value*, of a stable setoid $Y$ as a subsetoid of predicates on $Y$ that holds for at most one value and which is also stable. We denote the stable setoid of weak partial values over $Y$ as $\mathfrak{P}Y$:

$$\mathfrak{P}Y := \Sigma \mathcal{Y}\colon \wp Y.(\forall y. \neg\neg y \in \mathcal{Y} \Rightarrow y \in \mathcal{Y}) \land (\forall y_1\, y_2. y_1 \in \mathcal{Y} \Rightarrow y_2 \in \mathcal{Y} \Rightarrow y_1 \asymp y_2)$$

We define a *classical value*, also called a *weak value*, of stable setoid $Y$ as a subsetoid of $\mathfrak{P}Y$ that holds for at least one value. We denote this stable setoid of weak values over $Y$ as $\mathfrak{V}Y$.

$$\mathfrak{V}Y := \Sigma \mathcal{Y}\colon \mathfrak{P}Y. \check\exists y. y \in \mathcal{Y}$$

We use the weak existential here because we want to define functions for classical mathematics. If we had used the constructive existential, then our definition would end up almost equivalent to a constructive function.

Now a weak (partial) function can simply be defined as a function that produces weak (partial) values. Once we show that weak (partial) values form a monad (see Section 6.1), then we will see that a weak (partial) function is simply a Kleisli function for the weak (partial) value monad.

Has this shuffling around of definitions really helped us? The answer is yes, because both $\mathfrak{P}$ and $\mathfrak{V}$ are monads (see Section 6.1), users can use standard monadic combinators and programing techniques to manipulate functions and values of this type. Consider the example from before of trying to state $F(Gx) \asymp G(Fx)$. Using the the weak value monad, the functions $F$ and $G$ will have type $X \to \mathfrak{V}X$. Using the bind operation we can write the statement as $\check F(Gx) \asymp \check G(Fx)$. This statement is almost as easy to write as the classical notation, and certainly much easier than the long winded definition using binary predicates.

## 6.1 Monadic operations

Recalling our notation from Section 3.2, we define $\text{unit}_{\mathfrak{P}}\colon X \to \mathfrak{P}X$ as

$$\text{unit}_{\mathfrak{P}}\, x_0 := \{x\colon X \mid x \asymp x_0\}.$$

Given a function $f\colon X \to Y$, we define $\mathrm{map}_{\mathfrak{P}} f\colon \mathfrak{P} X \to \mathfrak{P} Y$ as

$$\mathrm{map}_{\mathfrak{P}} f \mathcal{X} := \{y\colon Y \mid \exists x \in \mathcal{X}. fx \asymp y\}.$$

Finally, we define $\mathrm{join}_{\mathfrak{P}}\colon \mathfrak{P}(\mathfrak{P} X) \to \mathfrak{P} X$ as

$$\mathrm{join}_{\mathfrak{P}} \mathcal{X} := \mathcal{X} \circ \mathrm{unit}_{\mathfrak{P}}.$$

The definitions of $\mathrm{map}_{\mathfrak{V}}$, $\mathrm{join}_{\mathfrak{V}}$, and $\mathrm{unit}_{\mathfrak{V}}$ are defined similarly.

Recall from Section 2.1 that we defined a function from booleans to propositions, $\lambda b.\langle b\rangle : \mathbb{B} \to \star$. There is a similar function taking classical boolean values to propositions. For $b\colon \mathfrak{V}\mathbb{B}$, we use the same notation to transform it into a proposition and context will make it clear which one is meant:

$$\langle b \rangle := \mathsf{T} \in b$$

## 6.2 Choice functions

The constructive axiom of choice is stated using the constructive quantifiers as follows:

$$\forall A B\colon \star. (\forall x\colon A. \Sigma y\colon B. R x y) \Rightarrow (\Sigma f\colon A \Rightarrow B. \forall x\colon A. R x (fx))$$

This is a theorem of dependent type theory (e.g. it is provable in MLTT [Nordström etal., 1990] and CIC [The Coq Development Team, 2009] and OTT [Altenkirch and McBride, 2006]) and I will call this the "constructive theorem of choice". Notice, however, that the function $f$ will not in general respect the equivalence relations between $A$ and $B$ if $A$ and $B$ are setoids.

Using the classical connectives and classical functions we can state the classical axiom of choice for setoids $X$ and $Y$ and a respectful relation $R$:

$$(\forall x y. \neg\neg R x y \Rightarrow R x y) \Rightarrow (\forall x\colon X. \exists y\colon Y. R x y) \Rightarrow (\exists f\colon X \to \mathfrak{V} Y. \forall x\colon X. fx \subseteq Rx)$$

This classical choice axiom is double negation stable by construction, and it is not a theorem of dependent type theory (e.g. it is not a theorem of MLTT nor CIC nor OTT). This is one significant difference from the usual approach of doing classical mathematics in a constructive logic by adding additional classical axioms. In the usual approach, the axiom of choice is inherited from the constructive theorem of choice.

However, it is worth noting that the classical axiom of choice is a consistent negative statement. In OTT, one can add consistent axioms to propositions without destroying canonicity of the non-propositional fragment [Altenkirch etal., 2007]. Thus in OTT, users have the flexibility to choose whether or not to include the axiom of choice in their work.

### 6.2.1 Definite choice

Although the general axiom of choice is not available without the use of other axioms, the principle of definite choice holds.

$$(\forall x y. \neg\neg R x y \Rightarrow R x y) \Rightarrow (\forall x\colon X. \exists! y\colon Y. R x y) \Rightarrow (\Sigma f\colon A \to \mathfrak{V}(B). \forall x\colon A. fx \subseteq Rx)$$

Notice that the existence of the classical function is even constructive.

The proof is pretty trivial because the hypothesis of definite choice contains exactly the ingredients needed to build the weak function $f$. Given an arbitrary $x$, we have that $Rx$ is a stable predicate, and there classically exists a unique $y$ such that $y \in Rx$ holds. Thus we can create a function $f\colon A \Rightarrow \mathfrak{V} B$ such that $\forall x. A. fx \subseteq Rx$ holds. Because $R$ is respectful, we can prove this function is respectful as well.

## 6.3 Examples of classical functions

One example of a classical function is the characteristic of a ring structure. For every ring $R$ there is a monoid homomorphism from $\rho\colon \mathbb{N} \to R$ defined by $\rho n := n \cdot 1_R$. The characteristic of $R$ is the GCD of the set $\{n \mid \rho n \asymp 0_R\}$.

In general, the characteristic of a ring is uncomputable, even if the ring enjoys decidable equality. However, we can create a classical function for it. More generally, we can create a classical function to compute the GCD of any subset of $\mathbb{N}$. First, we give the specification for the GCD:

$$\begin{aligned} \text{gcdSpec} &: \wp\mathbb{N} \Rightarrow \mathbb{N} \Rightarrow \star \\ \text{gcdSpec}\, P\, x &:= (\forall y \in P.\, \langle x|y\rangle) \wedge (\forall z.(\forall y \in P.\, \langle z|y\rangle) \Rightarrow \langle z\,|\,x\rangle) \end{aligned}$$

Notice that gcdSpec $P x$ is a stable proposition because $x|y$ is a decidable relation. We can also find a proof object for the proposition that there is a unique number $n$ satisfying gcdSpec $P$.

$$\text{gcdExists} \;:\; \forall P\colon \wp\mathbb{N}.\, \exists! n\colon \mathbb{N}.\, \text{gcdSpec}\, P\, n.$$

Because we have used the classical existential here, the entire theorem lies in the classical fragment of our logic. Therefore we can complete the proof using any of the usual classical techniques. Using definite choice, we can define the $\text{gcd}\colon \wp\mathbb{N} \to \mathfrak{V}\mathbb{N}$ operation.

We can prove various lemmas about this classical function. For example,

$$\forall PQ.\, \langle \text{gcd}\,(P \cup Q) \mathrel{\bar{|}} \text{gcd}\, P \rangle.$$

It is now easy to define the characteristic of a ring:

$$\text{char}\, R := \text{gcd}\, \{n \,|\, \rho(n) \asymp_R 0\}$$

Another example of a classical function is the degree function for polynomials. When the ring of coefficients enjoys decidable equality then the degree function is a fine constructive function. However, when decidable equality does not hold for the ring structure, as is the case for the real numbers, the degree function can no longer be constructed. However, even in this case, it is still important to be able to talk about bounds on degrees of polynomials in theorems.

In *A Course in Constructive Algebra* [Mines etal., 1988] the authors write (pg. 60–61),

> For $n \in \mathbb{N}$, a polynomial $f$ in $R[X]$ that can be written as $\sum_{i=0}^{n-1} r_i X^i$ is said to have *degree at most* $n-1$, written $\deg f \leqslant n-1$ or $\deg f < n$. [...] If $r_i \neq 0$ for some $i \geqslant d$, then we say that $f$ has *degree at least* $d$, and write $\deg f \geqslant d$. If $\deg f \leqslant d$ and $\deg f \geqslant d$ then we say that $f$ has *degree* d, written $\deg f = d$, [...]
>
> If $f$ and $g$ are polynomials, then we write $\deg f \leqslant \deg g$ if $\deg g < n$ implies $\deg f < n$ for each $n \in \mathbb{N}$; and we write $\deg f < \deg g$ if $\deg g \leqslant n+1$ implies $\deg f < n$ for each $n \in \mathbb{N}$. [...]

Notice all the effort that needs to go in defining all the possible ways of using and comparing degrees of polynomials. There are so many, it is unreasonable for the authors to write them all out. Indeed, it only takes them five lines before they write

> [...] Then $a\,b$ is the leading coefficient of $fg$, and $\deg fg = \deg f + \deg g$.

However, they have not defined what $\deg h = \deg f + \deg g$ means, although the reader can make up a reasonable definition on the spot.

Instead, classical functions can be used to give a rigorous definition of degree. We can define a specification for degree in a similar way to how we specified the GCD.

$$\text{degSpec}\, f\, d := \left( \sum_{i=0}^{d} r_i X^i \asymp f \right) \wedge \left( \forall n.\, \sum_{i=0}^{n} r_i X^i \asymp f \Rightarrow \langle d \leqslant n \rangle \right)$$

This specification is stable if the underlying ring $R$ is a stable setoid. In particular, this is the case for $\mathbb{R}$.

We can prove that for every polynomial $f$, there classically exists a unique $d$ such that degSpec $f d$ holds. From this proof we can create a weak function $\deg\colon R[X] \to \mathfrak{V}\mathbb{N}$. Using our monodic combinators, we can construct arbitrary propositions about degrees such as,

$$\begin{aligned} &\langle \deg f \;\bar{\leqslant}\; \hat{n} \rangle \\ &\langle \deg f \;\bar{\leqslant}\; \deg g \rangle \\ &\langle \deg f \;\bar{\equiv}\; \deg g \mathbin{\bar{+}} \deg h \rangle. \end{aligned}$$

This technique can express most of the propositions about degrees required, but there is a caveat. *A Course in Constructive Algebra* defines deg $f \geqslant n$ using constructive existentials. Therefore, there is no way to replicate this statement in this framework. However, I believe that the use of such constructive statements about degrees is relatively rare. In those cases where a constructive statement is required, then one would need to revert to the old method of making an ad-hoc definition.

## 7 Conclusion

One advantage of founding an interactive theorem prover on dependent type theory is that it has the capability of supporting both constructive reasoning and classical reasoning. However, one need not add axioms that destroy canonicity to get classical reasoning. Classical reasoning was already there because it is a fragment of constructive reasoning. This allows one to retain the benefits of constructive logic while still being able to derive classical results. We saw three examples of how to integrate traditionally classical results with constructive mathematics.

We also created a monadic method of defining classical functions spaces, again without the need to add classical axioms. We saw two examples of classical functions that can be used in constructive mathematics.

Maintaining canonicity in dependent type theory is important because it allows for computation to be done inside the logic using a technique known as reflection [Barendregt and Geuvers, 2001]. It also allows for program extraction, which means software can be developed along with their proofs of correctness. Dependent pair types (i.e. constructive existentials) are important in the development of correct software because it allows one to tie invariants right into data types [Leroy, 2009].

Prototype Coq modules for the weak connectives and weak value monads are given in Appendix A and Appendix B respectively. The GCD example has been formalized in Coq using these modules and the lemma that $\forall PQ. \langle \gcd(P \cup Q) \bar{\mid} \gcd P \rangle$ has been verified in order to show this proof of concept. Although the monadic functions work well for creating statements using classical functions, more theory of weak values needs to be developed and included in the module in order to make reasoning about them easier.

I hope that the idea of embedding classical reasoning this way will be developed further for proof assistants. While we have seen that it is possible to do classical reasoning, more support in constructive proof assistants would be needed if classically minded users are to be able to use them transparently for classical reasoning. With enough support, it should be possible for users to not even know that the underlying logic of the system is constructive. This would benefit constructive users as well, because they would be able to use classical results in their work.

# Appendix A   Classical Connectives for Coq

```
Require Import Coq.Classes.Morphisms.
Require Import Coq.Program.Basics.

Lemma forall_stable : forall A (P : A -> Prop),
 (forall a, ~~P a -> P a) -> ~~(forall a, P a) -> forall a, P a.
Proof.
firstorder.
Qed.

Lemma imp_stable : forall P (Q : Prop), (~~Q -> Q) -> ~~(P -> Q) -> P -> Q.
Proof.
firstorder.
Qed.

Lemma and_stable : forall (P Q : Prop),
 (~~P -> P) -> (~~Q -> Q) -> ~~(P /\ Q) -> P /\ Q.
Proof.
firstorder.
Qed.

Lemma iff_stable : forall (P Q : Prop),
 (~~P -> P) -> (~~Q -> Q) -> (~~(P <-> Q) -> (P <-> Q)).
Proof.
firstorder.
Qed.

Lemma not_stable : forall (P : Prop), ~~~P -> ~P.
Proof.
firstorder.
Qed.

Hint Resolve and_stable iff_stable not_stable : stable.

Hint Extern 3 =>  match goal with
   [|- (~ (~ (?P -> ?Q))) -> ?P -> ?Q] => apply (imp_stable P Q)
  |[H:(~ (~ (?P -> ?Q))) |- ?P -> ?Q] => apply (imp_stable P Q);[|apply H]
 end : stable.

Hint Extern 2 =>
 match goal with
   [|- (~ (~ forall d, @?Q d)) -> forall d, @?Q d] =>
     change (~~(forall d, Q d) -> forall d, Q d); apply forall_stable
  |[H:(~ (~ forall d, @?Q d))|- forall d, @?Q d] =>
     change (forall d, Q d); apply forall_stable;[|apply H]
 end : stable.

Ltac solveStable := solve [auto with stable |firstorder with stable].

Section OrW.

Variables (P Q : Prop).

Definition orW : Prop := ~(~P/\~Q).

Lemma orW_stable : ~~orW -> orW.
Proof.
firstorder.
Qed.
```

```
Lemma orWeaken : P \/ Q -> orW.
Proof.
firstorder.
Qed.

Lemma orW_elim : forall (R : Prop),
 (P -> R) -> (Q -> R) -> (~~R -> R) -> orW -> R.
Proof.
firstorder.
Qed.

End OrW.

Hint Resolve orW_stable : stable.
Hint Resolve orWeaken : core.

Ltac leftW := apply orWeaken; left.
Ltac rightW := apply orWeaken; right.
Tactic Notation "orWelim" constr(H) "as"
  simple_intropattern(A) simple_intropattern(B):=
  let G := fresh "orWelim" in
 destruct H as [A|B|G] using orW_elim;[| |solveStable].
Tactic Notation "orWelim" constr(H):=
  let G := fresh "orWelim" in
 destruct H as [H|H|G] using orW_elim;[| |solveStable].

Instance orW_imp_morphism : Morphism (impl ==> impl ==> impl) orW.
firstorder.
Qed.

Instance orW_inverse_imp_morphism :
  Morphism (inverse impl ==> inverse impl ==> inverse impl) orW.
firstorder.
Qed.

Instance orW_iff_morphism :
  Morphism (iff ==> iff ==> iff) orW.
firstorder.
Qed.

Lemma excluded_middle : forall P, (orW P (~P)).
firstorder.
Qed.

Section ExistsW.

Variables (A : Type) (P : A -> Prop).

Definition exW : Prop := ~forall a,~(P a).

Lemma existsW_stable : ~~exW -> exW.
Proof.
firstorder.
Qed.

Lemma existsWeaken : { a : A | P a } -> exW.
Proof.
firstorder.
Qed.

Lemma existsW_elim :
 forall (Q:Prop), (forall a, P a -> Q) -> (~~Q -> Q) -> exW -> Q.
Proof.
firstorder.
Qed.

End ExistsW.

Hint Resolve existsW_stable : stable.

Ltac existW x := apply existsWeaken; exists x.
Tactic Notation "existWelim" constr(H) "as"
  simple_intropattern(a) simple_intropattern(Ha):=
  let G := fresh "existWelim" in
 destruct H as [a Ha|G] using existsW_elim;[|solveStable].

Instance exW_imp_morphism {A : Type} :
  Morphism (pointwise_relation A impl ==> impl) (@exW A).
firstorder.
Qed.

Instance exW_inverse_imp_morphism {A : Type} :
  Morphism (pointwise_relation A (inverse impl) ==> (inverse impl))
           (@exW A).
firstorder.
Qed.

Instance exW_iff_morphism {A : Type} :
  Morphism (pointwise_relation A iff ==> iff) (@exW A).
firstorder.
Qed.

Notation "'existsW' x , p" := (exW _ (fun x => p))
  (at level 200, x ident, right associativity) : type_scope.
Notation "'existsW' x : t , p" := (exW _ (fun x:t => p))
  (at level 200, x ident, right associativity,
      format "'[' 'existsW'  '/ ' x : t ,  '/ ' p ']'")
  : type_scope.
```

# Appendix B  Classical Values for Coq

```
Require Import ClassicalConnectives.
Require Import Setoid.
Require Import Relation_Definitions.
Require Import SetoidClass.

Record ClassicSetoid : Type :=
 {CScarrier :> Type
 ;CSetoid : Setoid CScarrier
 ;CSstable : forall x y, ~~(x == y) -> (x == y)
 }.

Hint Resolve CSstable : stable.

Definition Function (A B:ClassicSetoid) : ClassicSetoid.
intros A B.
apply (Build_ClassicSetoid
 _ (Build_Setoid
  (@Equivalence.respecting_equiv _ _ (@setoid_equiv _ (CSetoid A))
                                 _ _ (@setoid_equiv _ (CSetoid B))))).
abstract (intros [f Hf] [g Hg] Hfg; simpl in *; auto with stable).
Defined.

Definition FunSpace (A B : ClassicSetoid) : Type := Function A B.

Definition FunctionApply A B (f:FunSpace A B) : A -> B := proj1_sig f.

Coercion FunctionApply : FunSpace >-> Funclass.

Instance CSetoid_ (A:ClassicSetoid) : Setoid A :=
 {equiv := @equiv _ (CSetoid A)
 ;setoid_equiv := @setoid_equiv _ (CSetoid A)
 }.

Add Parametric Morphism A B : (FunctionApply A B)
  with signature (equiv ==> equiv ==> equiv) as FunctionApply_morph.
Proof.
intros f g Hfg x y Hxy.
apply Hfg.
auto.
Qed.

Record WeakPartialValue (A:ClassicSetoid) :=
 {WPVcarrier : A -> Prop
 ;WPVstable : forall a, ~~WPVcarrier a -> WPVcarrier a
 ;WPVmorph : Morphism (@equiv A (CSetoid A) ==> iff) WPVcarrier
 ;WPVuniq  : forall x y, WPVcarrier x -> WPVcarrier y -> x == y
 }.

Hint Resolve WPVstable : stable.

Notation "wpv 'holds' x" := (WPVcarrier _ wpv x) (at level 70).

Record WeakValue (A:ClassicSetoid) := {
 WVcarrier :> WeakPartialValue A;
 WVexists : existsW x, WVcarrier holds x
 }.

(* equivalence relation for WeakValues and WeakPartialValues *)
Definition WPVeq (A:ClassicSetoid) : relation (WeakPartialValue A) :=
 fun x y => respectful (@equiv A (CSetoid A)) iff (WPVcarrier _ x)
                                                 (WPVcarrier _ y).

Lemma WPVeq_is_Setoid : forall A, Equivalence (WPVeq A).
Proof.
intros A.
split.
  intros [p] x y Hxy.
  rewrite Hxy.
  reflexivity.
 intros [p] [q] Hpq x y Hxy.
 symmetry; apply Hpq; symmetry.
 auto.
intros [p] [q] [r] Hpq Hqr x y Hxy.
transitivity (q y).
 apply Hpq; auto.
apply Hqr; reflexivity.
Qed.

Instance WPVSetoid (A:ClassicSetoid) : Setoid (WeakPartialValue A) :=
 {equiv := WPVeq A
 ;setoid_equiv := @WPVeq_is_Setoid A
 }.

Add Parametric Morphism A : (WPVcarrier A)
  with signature (equiv ==> equiv ==> iff) as WPVcarrier_morph.
Proof.
intros x y Hxy.
apply Hxy.
Qed.

Lemma WPVeq_stable : forall A (x y : WeakPartialValue A),
 ~~(x == y) -> x == y.
Proof.
intros A p q Hpq x y Hxy.
rewrite (WPVmorph _ p _ _ Hxy).
apply iff_stable; auto with stable.
revert Hpq.
cut (p == q -> (p holds y <-> q holds y)).
 tauto.
intros Hpq.
rewrite Hpq.
```

```
    reflexivity.
Qed.

Canonical Structure WPVCsetoid A :=
 Build_ClassicSetoid (WeakPartialValue A) (WPVSetoid A) (WPVeq_stable A).

Lemma WVeq_is_Setoid : forall A, @Equivalence (WeakValue A) (WPVeq A).
Proof.
intros A.
destruct (WPVeq_is_Setoid A).
split; eauto.
Qed.

Instance WVSetoid (A:ClassicSetoid) : Setoid (WeakValue A) :=
 {equiv := WPVeq A
 ;setoid_equiv := @WVeq_is_Setoid A
 }.

Canonical Structure WVCsetoid A :=
 Build_ClassicSetoid (WeakValue A) (WVSetoid A)
                     (fun x y => WPVeq_stable A x y).

Add Parametric Morphism A : (WVcarrier A) with signature (equiv ==> equiv)
    as WVcarrier_morph.
Proof.
auto.
Qed.

Definition WPVzero : forall A, WeakPartialValue A.
intros A.
exists (fun x => False).
  abstract (auto with stable).
 abstract (intros _ _ _; reflexivity).
abstract contradiction.
Defined.

Definition WPVunit : forall A:ClassicSetoid, A -> WeakPartialValue A.
intros A a.
exists (fun x => a == x).
  abstract (apply CSstable).
 abstract (intros x y Hxy; rewrite Hxy; reflexivity).
abstract (intros x y Hax; rewrite Hax; auto).
Defined.

Add Parametric Morphism A : (WPVunit A) with signature (equiv ==> equiv)
    as WPVunit_morph.
Proof.
intros x y Hxy.
intros a b Hab.
simpl.
rewrite Hxy, Hab.
reflexivity.
Qed.

Definition WVunit : forall A:ClassicSetoid, A -> WeakValue A.
intros A a.
exists (WPVunit A a).
abstract (existW a; simpl; reflexivity).
Defined.

Add Parametric Morphism A : (WVunit A) with signature (equiv ==> equiv)
    as WVunit_morph.
apply WPVunit_morph.
Qed.

Definition WPVmap : forall (A B : ClassicSetoid)
  (f : Function A B),
 WeakPartialValue A ->
 WeakPartialValue B.
intros A B f a.
exists (fun y => existsW x, a holds x /\ proj1_sig f x == y).
  abstract (auto with stable).
 abstract (intros x y Hxy;setoid_rewrite Hxy;reflexivity).
abstract(
intros x y Hx Hy;
existWelim Hx as x0 [Hx0 Hfx0];
existWelim Hy as x1 [Hx1 Hfx1];
rewrite <- Hfx0, <- Hfx1;
apply (proj2_sig f);
eapply WPVuniq; [apply Hx0 |apply Hx1]
).
Defined.

Add Parametric Morphism A B : (WPVmap A B)
    with signature (equiv ==> equiv ==> equiv) as WPVmap_morph.
Proof.
intros [f fmorph] [g gmorph] Hfg.
intros x y Hxy.
intros a b Hab.
simpl in *.
assert (Hfga : forall a, f a == g a).
 intros a0; apply Hfg.
 reflexivity.
setoid_rewrite Hfga.
setoid_rewrite Hxy.
setoid_rewrite Hab.
reflexivity.
Qed.

Definition WVmap : forall (A B : ClassicSetoid)
  (f : Function A B),
 WeakValue A ->
 WeakValue B.
intros A B f a.
exists (WPVmap A B f a).
abstract (
existWelim (WVexists _ a) as a0 Ha0;
existW (proj1_sig f a0);
simpl;
existW a0;
auto with *
```

```
).
Defined.

Add Parametric Morphism A B : (WVmap A B)
    with signature (equiv ==> equiv ==> equiv) as WVmap_morph.
Proof.
simpl.
intros f g Hfg.
intros x y Hxy.
apply WPVmap_morph;
auto.
Qed.

Definition WPVjoin : forall A,
  WeakPartialValue (WPVCsetoid A) -> WeakPartialValue A.
intros A a.
exists (fun x => a holds (WPVunit A x)).
  abstract (auto with stable).
 abstract (intros x y Hxy; rewrite Hxy; reflexivity).
abstract (
intros x y Hx Hy;
assert (Hyy : y == y);[reflexivity|];
assert (Hxy := WPVuniq _ _ _ _ Hx Hy y y Hyy);
simpl in Hxy;
rewrite Hxy;
reflexivity
).
Defined.

Add Parametric Morphism A : (WPVjoin A) with signature (equiv ==> equiv)
    as WPVjoin_morph.
Proof.
intros x y Hxy.
intros a b Hab.
simpl.
rewrite Hxy, Hab.
reflexivity.
Qed.

Definition WVjoin : forall A, WeakValue (WVCsetoid A) -> WeakValue A.
intros A a.
exists (WPVjoin _ (WPVmap _ _ (exist _ _ (WVcarrier_morph A)) a)).
abstract (
existWelim (WVexists _ a) as a0 Ha0;
existWelim (WVexists _ a0) as a1 Ha1;
existW a1;
simpl;
existW a0;
split; auto;
intros x y Hxy;
rewrite Hxy;
simpl; split;
[apply WPVuniq | intros Hy; rewrite <- Hy];
auto).
Defined.

Add Parametric Morphism A : (WVjoin A) with signature (equiv ==> equiv)
    as WVjoin_morph.
intros x y Hxy.
intros a b Hab.
simpl.
setoid_rewrite Hxy.
setoid_rewrite Hab.
reflexivity.
Qed.

Definition WPVBind (A B : ClassicSetoid)
 (f : Function A (WPVCsetoid B))
 (a : WeakPartialValue A) : WeakPartialValue B :=
 WPVjoin _ (WPVmap _ _ f a).

Definition WVBind (A B : ClassicSetoid)
 (f : Function A (WVCsetoid B))
 (a : WeakValue A) : WeakValue B :=
 WVjoin _ (WVmap _ _ f a).

Lemma WPVlaw1 : forall (A B : ClassicSetoid)
 (f : FunSpace _ _) a, WPVBind A B f (WPVunit A a) == f a.
Proof.
intros A B [f morph] a.
change (Morphism (equiv ==> equiv)%signature f) in morph.
intros x y Hxy.
simpl.
setoid_rewrite Hxy.
assert (Hyy : y == y).
 reflexivity.
split.
 intros E.
 existWelim E as x0 [Hx0 Hfx0].
 rewrite Hx0.
 rewrite (Hfx0 y y Hyy).
 auto.
intros Hfa.
existW a.
split.
 reflexivity.
intros c d Hcd.
rewrite Hcd.
simpl.
split.
 apply WPVuniq; auto.
intros Hyd.
rewrite <- Hyd.
auto.
Qed.

Lemma WPVlaw2 : forall (A : ClassicSetoid) (a : WeakPartialValue A),
 WPVBind _ _ (exist _ _ (WPVunit_morph A)) a == a.
Proof.
intros A a.
intros x y Hxy.
```

```
    simpl.
    setoid_rewrite Hxy.
    simpl.
    split.
     intros E.
     existWelim E as x0 [Hx0 Hfx0].
     setoid_replace y with x0.
      auto.
     rewrite <- (Hfx0 x0 x0); simpl; reflexivity.
    intros Hay.
    existW y.
    split; auto.
    change (WPVunit A y == WPVunit A y).
    reflexivity.
   Qed.

   Section Law3.

   Variable (A B C : ClassicSetoid)
    (f : FunSpace A (WPVCsetoid B))
    (g : FunSpace B (WPVCsetoid C))
    (a : WeakPartialValue A).

   Lemma Law3Help : Morphism (equiv ==> equiv) (fun x => WPVBind B C g (f x)).
   Proof.
   intros x y Hxy.
   unfold WPVBind.
   setoid_rewrite Hxy.
   reflexivity.
   Qed.

   Lemma WPVlaw3 :
    WPVBind B C g (WPVBind A B f a) ==
    WPVBind A C (exist _ _ Law3Help) a.
   Proof.
   intros x y Hxy.
   setoid_rewrite Hxy.
   clear x Hxy.
   split.
    intros E.
    existWelim E as x [E Hx].
    existWelim E as z [Hz Hfz].
    simpl.
    existW z.
    split; auto.
    change ((WPVBind B C g (f z)) == (WPVunit C y)).
    setoid_rewrite <- Hx.
    unfold WPVBind.
    setoid_rewrite Hfz.
    apply WPVlaw1.
   intros E.
   existWelim E as z [Hz Hfz].
   simpl.
   assert (Hyy : y == y).
    reflexivity.
   rewrite <- (Hfz y y Hyy) in Hyy.
   simpl in Hyy.
   existWelim Hyy as x [Hx Hgx].
   existW x; split; auto.
   existW z; split; auto.
   intros c d Hcd.
   rewrite Hcd.
   simpl.
   split.
    apply WPVuniq; auto.
   intros Hxd; rewrite <- Hxd; auto.
   Qed.

   End Law3.

   Lemma WVlaw1 : forall (A B : ClassicSetoid)
    (f : FunSpace A (WVCsetoid B))
    (a : A), WVBind A B f (WVunit A a) == f a.
   Proof.
   intros A B f a.
   intros x y Hxy.
   assert (morph : Morphism (equiv ==> equiv)
                            (fun x => (f x : WPVCsetoid B))).
    intros c d Hcd.
    rewrite Hcd.
    reflexivity.
   assert (X := WPVlaw1 A B (exist _ _ morph) a x y Hxy).
   rewrite <- X.
   split.
    intros E.
    existWelim E as b [E Hb].
    existWelim E as c [Hc1 Hc2].
    simpl.
    existW c.
    change (b == (WVunit B x)) in Hb.
    rewrite <-Hc2 in Hb.
    split; auto.
   intros E.
   simpl.
   existW (WVunit B x).
   split; auto.
   change (WVunit B x == WVunit B x).
   reflexivity.
   Qed.

   Lemma WVlaw2 : forall (A : ClassicSetoid)
    (a : WeakValue A), WVBind _ _ (exist _ _ (WVunit_morph A)) a == a.
   Proof.
   intros A a.
   intros x y Hxy.
   rewrite <- (WPVlaw2 A a x y Hxy).
   split.
    intros E.
    existWelim E as b [E Hb].
    existWelim E as c [Hc1 Hc2].
    change (b == WVunit A x) in Hb.
    rewrite <- Hc2 in Hb.
    simpl.
    existW c.
    split; auto.
   intros E.
   simpl.
   existW (WVunit A x).
   split; auto.
   change (WVunit A x == WVunit A x).
   reflexivity.
   Qed.

   Lemma WVlaw3 : forall (A B C : ClassicSetoid)
    (f : FunSpace A (WPVCsetoid B))
    (g : FunSpace B (WPVCsetoid C))
    (a : WeakPartialValue A),
    WPVBind B C g (WPVBind A B f a) ==
    WPVBind A C (exist _ _
            (Law3Help A B C f g)) a.
   Proof.
   apply WPVlaw3.
   Qed.

   Section Ap.

   Variable (A B : ClassicSetoid) (f : WeakValue (Function A B))
            (a : WeakValue A).

   Lemma Ap1 : forall f0 : FunSpace A B,
    Morphism (equiv ==> equiv) (fun a0:A => f0 a0).
   Proof.
   intros [f0 Hf0].
   apply Hf0.
   Qed.

   Definition ApBody (f0 : Function A B) : WeakValue B :=
    WVmap _ _ (exist _ _ (Ap1 f0)) a.

   Lemma Ap2 : Morphism (equiv ==> equiv) ApBody.
   Proof.
   intros x y Hxy.
   unfold ApBody.
   apply WVmap_morph.
   apply Hxy.
   reflexivity.
   Qed.

   Definition Ap : WeakValue B :=
    WVBind _ _ (exist _ _ Ap2) f.

   End Ap.
```